\documentclass{article}
\usepackage{amsmath}

\begin{document}

\title{A CRITICAL APPROACH TO TOTAL AND PARTIAL DERIVATIVES}

\bigskip

\author{{\bf Andrew E. Chubykalo and Rolando Alvarado Flores}}
\maketitle

\begin{center}
{\rm Escuela de F\'{\i}sica, Universidad Aut\'onoma de
Zacatecas \\
Apartado Postal C-580,\\
Zacatecas 98068, ZAC., M\'exico\\
e-mails: andrew@logicnet.com.mx  and  andrew@ahobon.reduaz.mx}
\end{center}



\baselineskip 7mm

\bigskip

\begin{center}
Received January 10, 2002
\end{center}

\begin{abstract}
In this brief note we critically examine the process of
partial and  of total differentiation, showing some of the problems that
arise when we  relate both concepts. A way to solve all the problems is
proposed.
\end{abstract}

$$$$$$$$


\clearpage

\section{Introduction}

Our article is devoted to the discussion of the total derivative concept,
a general and frequently applied concept of  mathematical analysis.
Indeed,  the derivatives play a significant role in  modern physical
theories and are present in many basic physical laws.

Considering some of the  basic statements of  classical electrodynamics,
one of the authors  (A.E.Ch) paid attention to the fact of some
inaccuracy of applying in physics  the concept of the partial derivative
of the many variable function (see [1,2]).  L. Schwartz [3] warned
prudently against  this inaccuracy  (the text after Eq.(I,2;5) in [3]):
``{\it In a short narrative one  identifies sometimes $f_1$ and $f$,
saying, that  is the same function (sic.!),  represented with the
help of the variable $x_1$ instead of $x$. Such a simplification is very
dangerous and may result in very serious contradictions.}"\footnote{Here
$f = f(x)$ is the function determined in a set ${\cal E}$ with the
values in a set ${\cal F}$, $f_1 = f\circ u$, where $u$ is the mapping of
some set ${\cal E}_1$ into the set ${\cal E}:  x = u(x_1)$.  L.Schwartz
calls the function $f_1$  a foretype of the function $f$ under a
variable substitution.}

Such an inaccuracy and  its consequences in some problems of physics
inspired our interest  to the total derivative concept in the classical
analysis under the condition of   {\it double} dependence from a time
variable $t$:  implicit and  explicit ones.  Such a situation is
characteristic of many physical problems, first of all of  classical
mechanics (see for example,  [4], where the author even introduces a new
special term as ``whole-partial derivative") and  classical
electrodynamics (see, for instance, Section.  4 in [2] where it was
considered in detail).

As a matter of fact, a concept of the partial derivative is
habitually associated with a concept of  the function of  many variables,
but a concept of the total derivative must be associated with another
function, which is some restriction of the function of  many variables.
L.Schwartz  did not introduce an additional co-term though he
denoted repeatedly [3] the  corresponding moment connected with applying
of the total derivative concept.

We are interested in the most important point, namely, in an origination
of the total derivative concept.

Thus, let us consider functions $E$  which are determined
as
\begin{equation}
E [x_1(t),\ldots, x_{n-1}(t), t]  = \;_{{\rm def}}E[{\bf x}(t),t],
\end{equation}
\begin{equation}
E(x_1,\ldots,x_{n-1},t) = \;_{{\rm def}}E({\bf x}, t),
\end{equation}
thereby emphasizing the need to distinguish the different functions:
$E[\;]$ of one variable, $E(\;)$  is the function of $n$ variables.

These functions evaluated at  different points of the globally defined
manifold $O = R^{n-1}\times R$ are a source  of confusion when we try to
calculate total or partial derivatives, and can lead us  to write down
something meaningless.  We shall explain this in detail in the next
sections, using, to save writing, the notation introduced in (1) and  (2)
and a theoretical framework that will show that the problems and
distinctions treated in this note have been not treated before.

Usually the functions $E[\;]$  and $E(\;)$ represent the same
physical value, being  different functions in their mathematical
origination.

 Note that the authors neither in [2] nor in [4] do not
distinguish this kind of function\footnote{as well as G.M.Fichtengoltz,
who considers the case of double (explicit and implicit) dependence of
functions on two variables,  see [5], p. 388.}.  Ambiguities in the
``notation" for partial differentiation also have been remarked by Arnold
[6] p. 226 (p. 258 in English translation) without further development.

Therefore,  an  unallowable identification of the functions $E[\;]$  and
$E(\;)$  happens quite often.

For instance,  in the well-known physical formulae
$$
\frac{d}{dt}E  = ({\bf V}\cdot\nabla)E + \frac{\partial}{\partial t}E
$$
(here ${\bf V} = \frac{d{\bf x}}{dt}$) and
$$
\frac{df}{dt}   =   (H,f)  +  \frac{\partial f}{\partial t},
$$
where  $f$ is some dynamical value,  $(H,f)$ is the Poisson bracket, $H$
is the Hamilton function, the  full derivatives in the $lhs$ and the
partial derivative in the $rhs$  are applied to the {\it different}
functions:  there are the functions $E[\;]$ and  $f[\;]$ in the $lhs$, and
there are the functions $E(\;)$ and $f(\;)$ in the $rhs$ of the equation.

The difference between the functions:
$$
E [x_1(t),\ldots, x_{n-1}(t), t]  = \;_{{\rm def}}E[{\bf x}(t),t],\quad
E(x_1,\ldots,x_{n-1},t) = \;_{{\rm def}}E({\bf x}, t)
$$
is usually  not remarked in the literature, and for this reason we can
often write down meaningless symbols like:
\begin{equation}
\frac{\partial}{\partial t}E[{\bf x}(t), t],
\end{equation}
and
\begin{equation}
\frac{d}{dt}E({\bf x},t).
\end{equation}

The  symbols (3) and (4) are meaningless, because the
process denoted by the operator of {\it partial}  differentiation can be
applied only to functions of several {\it independent} variables and  $E
[{\bf x}(t), t]$ is not {\it such} a function.  Meanwhile, the operator of
{\it total} differentiation with respect to a given variable can  be
formally applied to functions of one variable only.  However,  we have a
well-known formula to relate both concepts:

\begin{equation}
\frac{d}{dt}E  = ({\bf V}\cdot\nabla)E + \frac{\partial}{\partial t}E
\end{equation} (here ${\bf V} = \frac{d{\bf x}}{dt}$).

Let us show that, in this form, Eq. (5) cannot be correct. What  is the
correct argument for the symbol $E$ on both sides? If we say that the
correct argument for both sides is $[{\bf x}(t), t]$ we get the chain of
symbols  (3), but in this case, the operator of a partial
differentiation would indicate that we must construct a new function in
the form $(\partial E/\partial t)$, hence we use the following procedure:

$$
\lim\limits_{\Delta t\rightarrow 0}\left\{\frac{E\left[{\bf x}(t) + \Delta
t\frac{d{\bf x}(t)}{dt},\; t+\Delta t\right] - E[{\bf x}(t), t]}{\Delta
t}\right\}.
$$
But this  is the definition of total
differentiation! Thus, the symbols of  total and of partial
differentiation denote the same process, therefore,  because $E$ is the
same function on both sides of the equation, we get:
$$
({\bf V}\cdot\nabla)E[{\bf x}(t), t]  = 0
$$
always.  But even if the procedure which we followed were correct
(which it is not, of course!),  this equation is not correct for $E$ as a
function of the functions ${\bf x}(t)$, because the partial
differentiation would involve increments of the functions ${\bf x}(t)$ in
the form ${\bf x}(t) + \Delta{\bf x}(t)$ and we do not know how we must
interpret this increment because we have two options: {\it either} $\Delta
{\bf x}(t) = {\bf x}(t) - {\bf x}^*(t)$, {\it or} $\Delta{\bf x}(t) =
{\bf x}(t) - {\bf x}(t^*)$.  Both are different processes because the
first one involves changes in the functional form of the functions ${\bf
x}(t)$, while the second involves changes in the position along the path
defined by ${\bf x} = {\bf x}(t)$ but preserving the same functional form.
Hence, it is clear that we have here different concepts.  If we remember
the definition of partial differentiation, we can see where the mistake is:
``{\it the symbol:  $\frac{\partial}{\partial t}E({\bf x}, t)$ means that
we take the variations of $t$ when the values of} {\bf x} {\it are
constant}". It means that we make the only change $t + \Delta t$ in the
function.  But this is only possible if the coordinates {\bf x} are
independent from $t$.  Hence, we can see that the correct argument cannot
be $[{\bf x}(t), t]$, because, as we have shown, this supposition leads to
the incorrect result (5). If we make the other supposition, that the
correct argument is $({\bf x}, t)$ we can get the same conclusion, i.e.,
equation (5).  Hence, {\it none of these suppositions is correct}.  What
is the solution, then?  Actually, in the
equation (5) we have two {\it different} functions: on the left hand side
we have the function $E[{\bf x}(t),t]$ defined on a {\it curve} in a
$n$-surface and on the right hand side we have the function $E({\bf x},
t)$ defined on the {\it all}  $n$-surface, which obviously are {\it
quite} different functions, while  we have a limiting procedure to get
a unification of concepts in the realm of functions of one variable.

\section{Theoretical framework}

We  shall begin describing the elements that we shall use in the
development of the problem's explanation.  The globally defined
coordinates of our manifold $O$ are given by $\langle {\bf x},t
\rangle$, we define a function $E : O \rightarrow \Re$ where $\Re$ is the
real line.  Hence the values of the function $E$ on any point $O$  are
given as $E({\bf x}, t)$.  However, we shall be interested in the
1-dimensional subsets of $O$, hence we denote any of them by  $T$. To
describe this set (a path) we must introduce a function of the form:
\begin{equation}
p(t) = \langle {\bf x}(t),t \rangle
\end{equation}

Otherwise we can introduce this function in the form:
\begin{equation}
p(s) = \langle {\bf x}(s),t(s)\rangle
\end{equation}

Parametrization (6) for the path is a special case of (7) when we choose
the function: $t(s) = s = t$.  We suppose that $T\subset  O$ is
1-dimensional, hence a path in $O$. This path can be the integral curve
of a set of ordinary differential equations  (ODE's), we mean, it defines
the orbit of a 1-parametric group action  over $O$. If this action is a
free action we get that for any  pair $p,q \in O$  there exists an
integral curve joining them when we define one of them as  an initial
value. In this way the paths cover the manifold $O$ defining  a foliation
by 1-dimensional sheets.  The whole previous construction is  better
understood if we introduce the tangent vector space at each point  of $O$.
If the tangent vector field is defined at all the points of $O$ by the
equation:
\begin{equation}
\sum\limits_{i=1}^{n-1} f_i({\bf x}, t)\frac{\partial}{\partial x_i}
+ \frac{\partial}{\partial t},
\end{equation}
we can define the tangent vectors at each point of the path with the help
of the set of ODE's:
\begin{equation}
\frac{d x_i(t)}{dt} = f_i[{\bf x}(t), t];\qquad  i = 1 ,\ldots, n-1,
\end{equation}
whose  integral curves allow us to construct the 1-dimensional foliation
sheets.  The system (5) is the answer to a very important question that we
must keep in mind all along the work. The question is:
\begin{quotation}
(AA) How  can we construct sheets $T$ such that its tangent vectors are
given by the vector field with components $f_i$?
\end{quotation}

We  cannot overestimate the question. Hence our starting point is a
situation of total lack of knowledge regarding the form of the
1-dimensional sheet $T$. We just have the form of the vector field $f_i$,
that is, we know the distribution of vector fields in the space. In [7] we
have called this situation the \{$f$\}-case. In any usual treatment of the
subject of differential geometry   any  distribution of vector spaces if
the system (9) is solvable is called ``integrable".  Substantially the
same is done to construct foliation in more dimensions, the obvious change
is that we don't want 1-dimensional sheets, instead, we want to construct
$m < n -1$ dimensional submanifolds at each point. The basic question is,
again (AA).  We must remark something very important, when we pass from
(8) to (9) we have changed the functions $f_i (x, t)$ evaluated at any
point on the manifold $O$, by the functions $f_i (x(t), t)$ which are
evaluated on the sheets of the foliation. We have supposed, to do this
operation, that the sheets exist. To prove this supposition, it is usual
to impose a Holder condition on the vector fields, this is
sometimes enough to prove the existence of solutions by fixed point
arguments.  However, when we can construct solutions for the system, a
formal proof may be avoided.  This change from the globally defined
manifold to the  local integral sheets of  the foliation is done
noncritically. This criticism is what we shall carry out in the next
sections inside the framework described in this section.

\section{The problem}

We give the curve $T$ using the parametrization (6) but we write down:

\begin{equation}
 g (t) = E\circ p(t)
\end{equation}

where $g(t)$ is 1-variable function, $E$ and $p$ denote an $n$-variable
function defined on $O$ and a curve on $O$, respectively.  This way of
writing down the functions involved is more precise than the usual
notations (1) and (2). For this reason only   the use
of notations like (1) and (2) should be suppressed.  The really important
task becomes apparent when we try to differentiate totally or partially
the functions (1) and (2). If we want to partially differentiate (2) we
have no problem, because the usual definition of a partial derivative
requires that {\it we must change one of the variables while keeping
the rest constant}.  In the same way, if we want to derive with respect to
$t$ the expression (1)  we should not have any problem, because it is a
differentiation of a one variable function when we know the forms of $E$
and $p$. Let's show now the problems.

\begin{quotation}
(A) If we want to differentiate totally  (2) in any variable without
using, for the moment, any path, a moment  of reflection shows us that we
really employ the definition of a partial  derivative. Hence our use of the
symbol $dE({\bf x},t)/dt$ is wrong.  It is so because the
function is an n-variable function and our conventions  for the use of
symbols (the syntactical rules) tell us that for $n$-variable  functions
the correct notation is $\partial$. Hence, the symbol $dE({\bf x},t)/dt$ is
not correct, but it is very easy to write down the right symbols.
\end{quotation}

\begin{quotation}
(B) Now, if we want to partially differentiate (1) with respect to $t$, a
moment of  reflection shows us that when we try to give an increment to
$t$, while keeping  constant the other set of variables, this last
condition cannot be fulfilled  because if $t$ is incremented by an amount
$\Delta t$ the other variables are incremented  by an amount $(d{\bf
x}(t)/dt)\Delta t$. Hence we cannot keep them constant and we  have a
problem here, because we cannot apply the usual definition of partial
differentiation to expression (1).  \end{quotation}

To  our knowledge, this situation has not been pointed out before in the
usual treatises. However, a brief search in the corresponding literature
gives us the paper by K. Brownstein [4] where the concept of
so called ``whole partial derivative" is introduced. Let's see if it gives
us some light.  We shall use the framework outlined in section 2.
Brownstein starts with a function of the form:

\begin{equation}
G[\Phi(q_1, q_2, q_3, q_4), q_1, q_2, q_3, q_4],
\end{equation}
which falls within the  scope of our general  framework. To see this we
write down the following coordinate cover:
\begin{equation} \langle \Phi_1(q_1, q_2,
q_3, q_4), \Phi_2(q_1, q_2, q_3, q_4), \Phi_3(q_1, q_2, q_3, q_4), q_1,
q_2, q_3, q_4\rangle,
\end{equation}
which  is more general than the one
used by Brownstein in function (11).  In this case  we are over a
4-dimensional manifold represented on a 7-dimensional  manifold. We can
generate 4-dimensional sheets such that the manifold be  covered by them
if we can integrate the following system of first order coupled partial
differential equations:
\begin{equation} \frac{\partial\Phi({\bf
q})}{\partial q_j} = F_i(\Phi_1({\bf q}),\Phi_2({\bf q}), \Phi_3({\bf q}),
q1 ,q2 ,q3, q4);\qquad i = 1,\ldots,3 , j = 1,\ldots,4.
\end{equation}

The generated group of  transformations is a 4-parametric group. Just like
before, the functions $F_i$ are the components of a known vector field over
the manifold. And of course, here the same noncritical change has been
realized, because we have started at the manifold, and we have finalized
at its sub-manifold. A very important feature, which we shall express
later in the 1-dimensional case, appears here. The partial differential
equations (13) must be compatible differential equations, that is: their
cross derivatives must coincide. However, to establish the conditions we
must partially differentiate the functions at the right of (13), but this
is not possible because of the argument already used (B). The cross
differentiation condition leads us to the usual condition of commutativity
of vector fields, or in general grounds to the generators of a Lie
algebra, basic to the Frobenius' theory of integrability. However, we see
that we don't know how to calculate this condition because of the argument
(B).

Coming back to Brownstein's case let's discover again the same
difficulties as in the 1-dimensional. If we suppose, as Brownstein does,
that we can partially differentiate the function $G$ in (11) with respect
to any q variable, we fall again within the argument (B): the definition
of the partial derivative requires that we must change one of the $q$s
only.  But if we change any $q$ by an amount $\Delta q$ we change
the variable $\Phi$ by an amount $(\partial\Phi/\partial q)\Delta q$.
Hence, Brownstein's concept is arguable because we believe he makes  the
same mistakes that we are pointing out here.  Brownstein's mistake is as
follows: he must first define the partial derivatives which appear in {\it
his} formula (10) [4]\footnote{Eq. (10) in [4]:  $$ \frac{\bar\partial
G}{\bar\partial q^3}=\frac{\partial G}{\partial\Phi}
\frac{\partial\Phi}{\partial q^3} +\frac{\partial G}{\partial q^3}
$$}, derivatives which are in doubt because
of the argument (B).  So we can conclude that he achieves his goal: to
introduce a new symbol and a new name, but based on noncritical
concepts.

\section {A solution}

Let's  continue with our critical analysis. For this we shall write down a
highly  incorrect (because of the argument (B)), but nonetheless, very
popular, expression:
\begin{equation} \frac{dE[{\bf x}(t), t]}{dt} =
\sum\limits_{i=1}^{n-1}\frac{dx_i}{dt}
\frac{\partial E[{\bf x}(t), t]}{\partial x_i} + \frac{\partial
E[{\bf x}(t), t]}{\partial t}.
\end{equation}
One of the first mistakes is the following:  it is supposed that the
function $E[{\bf x}(t), t]$ is the same on both sides of the equation.
Let's apply the (B) argument to (14):
\begin{quotation}

1. At the right of the equation we see formations like $\partial E[{\bf
x}(t), t]/\partial t$, which by the use of the argument (B), have been
shown to be wrongly defined.

2. At  the left we see the symbol $dE[{\bf x}(t), t]/dt$ which is not
defined because  its ``definition", the right side, is wrongly defined,
and we have no other  definition for $dE[{\bf x}(t), t]/dt$. Hence, we
don't know how to calculate it.

\end{quotation}

Thus, a solution is required.  This can be obtained with the help of
equation (10) and some distinctions based upon it.  The first and most
important thing is to suppose that $g(t) = E\circ p(t)$ is a 1-variable
function only and that it is known. Hence, the usual definition of
1-variable derivation is available.  This supposition implies that we
must know the path $p$ and the functional form of $E$. We have analyzed
this supposition in detail in another paper [7], and so we shall
not repeat it.  Hence, it is the case that on the left side of (14)
 $dg(t)/dt$ must appear and not the function $E$ defined along the path.
On the right side the function $E[{\bf x}, t]$ must appear to get a
partial derivative using the usual definition. Finally, as it is the
case that the function $d{\bf x}(t)/dt$ is defined on one common point of
a class of paths and not all over the space $O$, as is the case for
$E({\bf x}, t)$ we shall write instead of  $d{\bf x}(t)/dt$ the functions
$f_i ({\bf x}, t)$ defined all over $O$ to get on the right hand side the
expression:
\begin{equation}
\sum\limits_{i=1}^{n-1}f_i({\bf x},t)
\frac{\partial E({\bf x}, t)}{\partial x_i} + \frac{\partial
E({\bf x}, t)}{\partial t}.
\end{equation}

But what is the relation between $dg(t)/dt$ with the expression  (15)?
We cannot make them equal all over the space $O$, because this is not
correct, we shall fall in previous mistakes again.  However both
expressions must be the same over a path, hence we  approximate the
expression (15) to the points of one path with the  help of a limiting
procedure:
\begin{equation}
\frac{dg(t)}{dt} = \lim\limits_{x\to p} \left(
\sum\limits_{i=1}^{n-1}f_i({\bf x},t) \frac{\partial E({\bf x},
t)}{\partial x_i} + \frac{\partial E({\bf x}, t)}{\partial t}\right)
\end{equation}

We have  discussed in detail several ways to use this expression in [7]
using the supposition that, in fact, we have integrability. Here we shall
just discuss the uses in the non-integrable case. But first let's remark
the advantages of (16):

\begin{quotation}
1.- On the left  hand side we have a function of just one variable, hence
the definition of derivative is clear.

2.- On the right  hand side we have only usual partial derivatives and
n-variable functions,  hence the usual definition of partial derivative is
clear.
\end{quotation}

The meaning of  the limiting procedure is very simple: on the globally
defined manifold $O$ we shall make that the variables that describe it
tend to the point of the path in some specified way. Of course this can be
done in many ways and depends on the topological properties of the
manifold. In any simple connected manifold the way in which we get the
points of the path should be not important.  The most common case of this
approximation procedure is the one which answer the question (AA), that is,
we approximate the tangent vectors of the path to the vectors given by the
vector field of components $f_i$.

\section{Some uses of the formula (16)}.

Let's show how the formula (16)  can be used in differential geometry. For
the sake of completeness we shall expose what is commonly considered as the
right procedure, and then we shall show that it can be done with our
methods, too.  Take an abstract manifold $M$ of dimension $N$ and define
over it a path $p$. Hence its coordinate representatives are given by:
\begin{equation}
\langle x_1[p(t)],\ldots, x_N[p(t)] \rangle.
\end{equation}

The usual goal  is to define in an intrinsic manner the tangent vectors,
that is, in such a way that they depend on the points of $M$ only and not
in the space in which $M$ is contained.  This can be done by defining the
tangent vectors in terms of the path $p$ in the following way: we define
the equivalence class of $p$ as
\begin{equation}
(p) = \;_{{\rm def}} \left\{p^*| \frac{dx_i[p(t)]}{dt}|_{t=0}
=\frac{dx_i[p^*(t)]}{dt}|_{t=0}\right\},
\end{equation}

in words: a path  $p$ is equivalent to a path $p^*$ if and only if at the
common point $p(0) = p^*(0)$ they have the same tangent vector. With this
definition it is a usual matter to prove that the  directional derivative
of the path, directed along the tangent vectors is  independent of the
selected path. To do this it is necessary to use   the chain rule to
write down the following two expressions:
\begin{equation}
\frac{df[p(t)]}{dt} =\sum\limits_{i=1}^{N}\frac{\partial f}
{\partial x_i}\,\frac{dx_i[p(t)]}{dt} \qquad {\rm and}\qquad
\frac{df[p^*(t)]}{dt}=\sum\limits_{i=1}^{N}\frac{\partial f}{\partial x_i}
\, \frac{dx_i[p^*(t)]}{dt}.  \end{equation}

We  can see that in the limit $t \to 0$ both expressions are the same,
hence the derivative is independent of the underlying path.  If we
remember the argument (B) we cannot  write equations
(19), hence, we must  use the correct expressions.  For that, we take two
paths which have only  one common point, hence we have two functions $g$
and $g^*$ equal in at least one  point. The expression at the right of
(16) under the sign of limit does not  depend on the path, and thus
remains the same, however the limit change  because the process of
approximation must be done considering two different paths, hence we write:
\begin{equation}
\frac{dg(t)}{dt} = \lim\limits_{x\to p}\sum\limits_{i=1}^{N}
V_i \frac{\partial f}{\partial x_i}\qquad {\rm and}\qquad
\frac{dg^*(t)}{dt} = \lim\limits_{x\to p^*}\sum\limits_{i=1}^{N}
V_i \frac{\partial f}{\partial x_i}.
\end{equation}

If we suppose that the tangents to the paths tend to the vector
field {\bf V}  we can write down the right hand sides of (20) as the right
hand sides of  (19). But this is equivalent to writing down:
\begin{equation}
\frac{dx_i[p(t)]}{dt} = V_i\{{\bf x}[p(t)]\}\qquad {\rm and}\qquad
\frac{dx_i[p^*(t)]}{dt} = V_i\{{\bf x}[p^*(t)]\}.
\end{equation}

Now,  if we suppose that as our initial value we have $p(0) = p^*(0) =
p_0$ we shall get the same path by invoking the usual theorems.  Hence the
underlying paths are not important and our process is well defined over
the equivalence classes.  But we have a bonus, when we cannot use the
equations (21), which is the case when our limiting procedure does not
converge  to the tangent vectors, that is, when we cannot find a
1-dimensional  foliation of the manifold by 1-dimensional integral paths
of the vector  field, we can express this condition in a quite simple
manner
\begin{equation}
\lim\limits_{x\to p} V_i(x_1,\ldots, x_N) \neq \frac{dx_i}{dt}
\end{equation}

Which,  of course, makes the integrability a part of the topological
properties of the manifold.

\section{On the integrability concept}

The  notion of integrability which we have reached in the previous
section must be compared with the most usual notion based on cross
differentiation, that is, the generation of a Lie algebra structure for
the generators of a distribution {\bf E} of vector spaces along any
manifold.  For  the case of differential 1-forms the conclusions about
integrability can be obtained with the use of its associated complex, the
De Rham complex, where the integrability condition of a differential
1-form $w$ can be expressed with the notation: $dw \neq 0$ because if it
is zero, by the use of the usual local Poincare's lemma we get a local
integral $f$ of the form $w = df$.  Another way to express the same
condition is with the use of the integral invariants of Cartan. For the
case of differential 1-forms we have that, if $f$ is such a scalar
invariant, hence the distribution of vector fields ${\bf E}$ along the
space is such that: ${\bf E} \subseteq {\tt Ker} (df)$ [8].  If we use
$n$-forms, $\varphi$, the condition is ${\bf E} \subseteq {\tt Ker}
(\varphi) \cap  {\tt Ker} (d\varphi)$ (Hence the $n$-form is an integral
invariant of Cartan).

However this notion is based  on one idea:  the cross differentiation of
expressions like the functions at the right hand side of (13) process
which we have showed to be meaningless because we don't have a way to
compute it. But with the development of the free coordinate tensor
formalism many things were left aside, and it was possible  to express the
conditions in free coordinate terms which avoid the explicit use of
operations like the usual total differentiation.  In this sense, we believe
that in the free coordinate tensor formalism   such problems
like the one treated in this note cannot be found,  except  in the cases
in which the total derivatives appear.  Coming back to the  notion of
integrability, even in the tensor formalism it is based on the  idea of
Lie algebra, which is a formal reconstruction of the idea of cross
differentiation, but the notion of integrability given by the equation
(22)  is not based on this notion as a primitive notion, instead it is
based on  the idea of appropriation to a given curve as a primitive
concept.

\section{Summary and discussion}

The brief treatment given here  suggests that  a profound, case
by case, investigation of the uses  of the formalism introduced in
differential geometry and topology is necessary.  However, probably the
usual tools must be complemented with a critical view  of the subject
involved, because  our representations of the underlying processes
may not be the same and this is the origin of the ignorance  of the
problem.  Really, the problem arises in the language, not in the usual
formalism, because we take seriously the idea that a partial  derivative
can be defined only when all the variables are constants  except one.
Trying to respect this definition is the source of everything.  This
definition defines two syntactical rules of the form:
\begin{equation}
d/dt : C(\Re,\Re^m)\rightarrow C(\Re,\Re^m); \qquad
\partial/\partial q : C(\Re^n,\Re^m)\rightarrow C(\Re^n,\Re^m)
\end{equation}
one for each operator. The functor $C$ should be  taken as adequate for
each case. The syntactical rules are, of course,  that the symbol $d/dt$
can only operate when the set represented by $\Re$ doesn't  appear as a
cartesian product, that is, its exponent can be only 1,  which means
1-variable function. For the operator of partial differentiation,  the set
must have a cartesian exponent different from 0 and 1, that is,  we
consider only $n$-variable functions $(n \neq 0, 1)$. Hence, if we look
more closely at the equation $g = E \circ p$ we can write down the
sequence:  $\Re \rightarrow \Re^n \rightarrow \Re^n$ which shows that any
derivative of the function $g$ must be a $d/dt$.  The composition
operation is of the syntactical form:
\begin{equation}
\circ: C(\Re^n, \Re^m) \times C(\Re, \Re^n) \rightarrow C(\Re, \Re^m),
\end{equation}
which shows that  its action produces a 1-variable function. Usually this
kind of rule is  not taken into account, and  people proceed with
heuristic arguments  based on one or other representation of the subject,
of even without  representation just by operating on the symbols. This is
not really wrong,  or at least that is our opinion. However, if one tries
to take  the  propositions seriously  a moment of reflection
over our own concepts  and the use which we made of them is necessary.
This is the heart of our attempt in this note.

The consistency of mathematical analysis it is the mayor
problem which motivated the new approaches. See, for example, very
interesting paper by K.  Brownstein [4], although we criticize some
aspects of this work. After our paper has  been already  submitted to this
Journal we have discovered a brilliant work by R.M. Santilli
``Nonlocal-integral isotopies of differential calculus, mechanics and
geometries" [9] where the author settles a new approach to differential
calculus (see, e.g., [9] p. 19, 1.5``Isodifferential calculus").

\begin{center}
\Large{{\bf Acknowledgments}}
\end{center}

The authors would like to express their gratitude to Prof. Valeri
Dvoeglazov for his discussions and critical comments. We would also like
to thank Annamaria D'Amore for revising the manuscript.  One of the
authors (RAF)  is in debt with the Centro de Estudios Multidisciplinarios
and  specially with its director Ing. Rogelio Cardenas Hernandez for his
constant support.

\end{document}